\documentclass[showpacs,showkeys,superscriptaddress,nofootinbib,floatfix,twocolumn,groupedaddress]{revtex4}
\usepackage{color}
\usepackage{amsfonts}
\usepackage{amsbsy}
\usepackage{mathrsfs}
\usepackage{graphicx}
\def\lsim{\mathrel{\rlap{
\lower4pt\hbox{\hskip-3pt$\sim$}}
    \raise1pt\hbox{$<$}}}     %less than approx. symbol
\def\gsim{\mathrel{\rlap{
\lower4pt\hbox{\hskip-3pt$\sim$}}
    \raise1pt\hbox{$>$}}}     %greater than or approx. symbol
\def\scr#1{\mbox{\scriptsize #1}}
\begin{document}
\title{Transverse-Mass Effective Temperature in Heavy-Ion Collisions
  from AGS to SPS}
%energies}
%
\author{Yu.B.~Ivanov}%\thanks{e-mail: Y.Ivanov@gsi.de}
\affiliation{Gesellschaft
 f\"ur Schwerionenforschung mbH, Planckstr. 1,
D-64291 Darmstadt, Germany}
\affiliation{Kurchatov Institute, Kurchatov
sq. 1, Moscow 123182, Russia}
\author{V.N.~Russkikh}%\thanks{e-mail: russ@ru.net}
\affiliation{Gesellschaft
 f\"ur Schwerionenforschung mbH, Planckstr. 1,
D-64291 Darmstadt, Germany}
\affiliation{Kurchatov Institute, Kurchatov
sq. 1, Moscow 123182, Russia}
\begin{abstract}
Transverse-mass spectra in Au+Au and Pb+Pb collisions in incident
energy range from 2$A$ to 160$A$ GeV are analyzed within the model of 3-fluid
dynamics (3FD). It is shown that dynamical description of freeze-out,
accepted in this model, naturally explains the  incident energy
behavior of inverse-slope parameters of these spectra observed in
experiment. 
Simultaneous reproduction of the inverse-slopes of all considered
particles ($p$, $\pi$ and $K$) %within the 3FD model 
suggests that these particles belong to 
the same hydrodynamic flow at the instant of their freeze-out. 
\end{abstract}
%
%\today

\maketitle

%\section{Introduction}

Experimental data on transverse-mass spectra of kaons produced in
central Au+Au \cite{E866} or Pb+Pb \cite{NA49} collisions reveal
peculiar dependence on the incident energy. The inverse-slope
parameter (so called effective temperature $T$) of these spectra at
mid rapidity increases with incident energy in the energy domain of
BNL Alternating Gradient Synchrotron (AGS) and then saturates at
energies of CERN Super Proton Synchrotron (SPS). In Refs.
\cite{Gorenstein03,Mohanty03} it was assumed that this saturation is
associated with the deconfinement phase transition. This assumption
was indirectly confirmed by the fact that microscopic transport
models, based on hadronic degrees of freedom, failed to reproduce
the observed behavior of the kaon inverse slope \cite{Bratkovskaya,Wagner}.
Hydrodynamic simulations of Ref. \cite{Hama04} succeeded to describe
this behavior. However, in order to reproduce it these hydrodynamic
simulations required incident-energy dependence of the freeze-out
temperature which almost repeated the shape of the corresponding
kaon effective temperature. This happened even in spite of using
equation of state (EoS) involving the phase transition into
quark-gluon plasma (QGP). This way, the puzzle of kaon effective
temperatures was just translated into a puzzle of freeze-out
temperatures. Moreover, results of Ref. \cite{Hama04} imply that
peculiar  incident-energy dependence of the kaon effective
temperature may be associated with dynamics of freeze-out.

In this paper we would like to present calculations of effective
temperatures within the
3FD model \cite{3FD,3FDm,3FDflow} which is suitable for
simulating heavy-ion collisions in the range from AGS to SPS energies.
%
%Unlike the conventional hydrodynamics, where local instantaneous
%stopping of projectile and target matter is assumed, a specific
%feature of the dynamic 3-fluid description is a finite stopping
%power resulting in a counter-streaming regime of leading
%baryon-rich matter.  The basic idea of a 3-fluid approximation to
%heavy-ion collisions \cite{3FD,3FDm} is that at each space-time
%point the generally nonequilibrium
%distribution function of baryon-rich
%matter, can be represented as a sum of two
%distinct locally equilibrated contributions,
%initially associated with constituent nucleons of the projectile
%(p) and target (t) nuclei. In addition, newly produced particles,
%populating the mid-rapidity region, are associated with a fireball
%(f) fluid.
%
We perform 
%have started 
our simulations \cite{3FD,3FDflow}
with a simple, hadronic EoS \cite{gasEOS} which involves only
a density dependent mean field providing saturation of cold
nuclear matter at normal nuclear density
%$n_0=$ 0.15 fm$^{-3}$
and
with the proper binding energy.
% -16 MeV.
%This EoS is a natural
%reference point for any other more elaborate EoS.
The 3FD model with the intermediate EoS  turned out to be able
to reasonably
reproduce a great body of experimental data \cite{3FD} in a wide
energy range from AGS to SPS. In particular, transverse-mass spectra
of protons were reproduced.
This was done with the unique set of
model parameters summarized in Ref. \cite{3FD}.
Problems were met only in description of transverse flow \cite{3FDflow}.
The directed flow requires a softer EoS at top AGS and SPS energies
(in particular, this desired softening may signal occurrence of the
phase transition into QGP). Similar softening is needed for
reproduction of recent data on rapidity distributions of net-baryon
number in central Pb+Pb collision at energies 20$A$--80$A$ GeV 
\cite{3FD-GSI07}. 
%The elliptic flow demands a more precise
%description of the nonequilibrium
%transverse-momentum anisotropy at the initial
%stage of nuclear collisions.

\begin{figure}[thb]
\includegraphics[width=8.5cm]{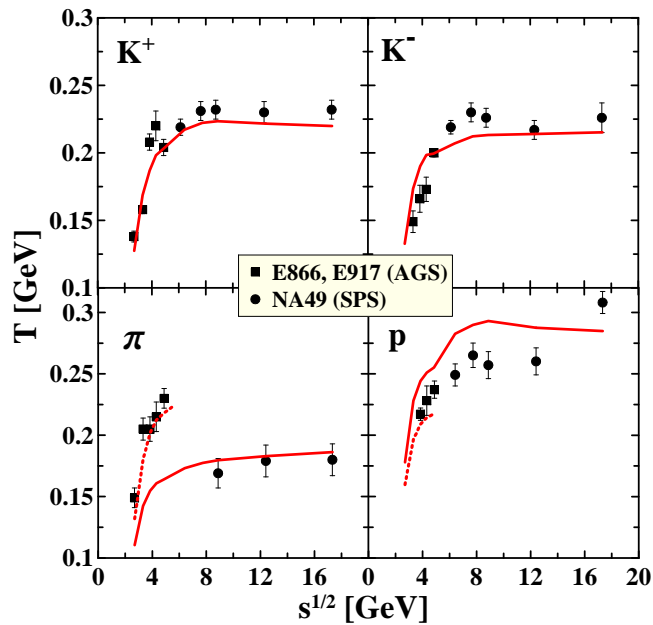}
\caption{(Color online)
Inverse-slope parameters of transverse-mass spectra of
kaons, pions and protons at mid rapidity produced in central
Au+Au  and  Pb+Pb collisions as a function of
invariant incident energy.
Solid lines correspond to  purely exponential fit
($\lambda=0$, see Eq. (\ref{Ttr})), 
while dashed lines present results with $\lambda=-1$
for pions and with $\lambda=1$ for protons.
Experimental data are from Refs. \cite{E866,NA49,E917p,NA49p}.
}
\label{fig1}
\end{figure}

The  transverse-mass spectra are most sensitive to the freeze-out
parameters of the model. In fact, inverse slopes of these spectra
represent a combined effect of the temperature and collective
transverse flow of expansion.
Fig. \ref{fig1} demonstrates these important interplay.
Had it been only the effect of thermal excitation, inverse slopes
for different
hadronic species would approximately equal.
%\footnote{Since kaons are
%predominately emitted from hotter parts of the system, their inverse
%slopes would be somewhat higher.}
%
The collective transverse flow makes them different.
These two effects partially
compensate each other: the later freeze-out occurs, the lower
temperature and the stronger collective flow are.
Nevertheless, transverse-mass spectra turn out to be sensitive
to the instant of the freeze-out.

3FD results for inverse-slope parameters of transverse-mass spectra
of kaons, pions and protons  produced in central Au+Au  and  Pb+Pb
collisions are presented in Fig. \ref{fig1}. The inverse slopes $T$
were deduced by fitting the calculated spectra by the formula
\begin{eqnarray}
\label{Ttr}
\frac{d^2 N}{m_T \; d m_T \; d y} \propto
\left( m_T\right)^\lambda
\exp \left(-\frac{m_T}{T}  \right),
\end{eqnarray}
where $m_T$ and $y$ are the transverse mass and rapidity, respectively.
%Solid lines in Fig. \ref{fig1} are calculated with
%$\lambda=0$ (purely exponential fit).
Though the purely exponential fit with $\lambda=0$
does not always provide the best fit of the spectra, it allows a
systematic way of comparing spectra at different incident energies.
In order to comply with experimental
fits at AGS energies (and hence with displayed experimental points),
we also present results with $\lambda=-1$ for pions
and with $\lambda=1$ for protons.
%which are displayed by dashed lines.
These results are obtained with precisely the same EoS and set of
parameters (friction, freeze-out and formation time) as those used in Ref.
\cite{3FD}, which was found to be the best for other observables.
No special tuning was done to reproduce these effective
temperatures.

Numerical problems, discussed in Ref. \cite{3FD}, prevented us from
simulations at RHIC energies. Already for the central Pb+Pb collision 
at the top SPS energy the code requires 7.5 GB of (RAM) memory.
% to be executed in reasonable time.
At the top
RHIC energy, required memory is three order of magnitude higher, which is
unavailable in modern computers.

As seen from Fig. \ref{fig1}, reproduction of effective temperatures
is quite reasonable. Moreover, the pion and proton effective
temperatures also reveal saturation at SPS energies, if they are
deduced from  the purely exponential fit with $\lambda=0$.
It is important that it is achieved with a single
freeze-out parameter $\varepsilon_{\scr{frz}}= 0.4$ GeV/fm$^3$,
the critical freeze-out energy density, which is the same for all
considered incident energies above 2$A$ GeV,
both for chemical and thermal freeze-out.
Only for smaller energies we used smaller values: 
$\varepsilon_{\scr{frz}}(2A \mbox{ GeV}) = 0.3$ GeV/fm$^3$ and 
$\varepsilon_{\scr{frz}}(1A \mbox{ GeV}) = 0.2$ GeV/fm$^3$. 
%\footnote{
%[In the main text of Ref. \cite{3FD} the value of approximately
%0.2 GeV/fm$^3$ was mentioned, which is the actual value of
%the freeze-out energy density ($\varepsilon_{\scr{out}}$, see below)
%rather than "trigger" value $\varepsilon_{\scr{frz}}$.]
%}.
In order to clarify why this happens, let us
turn to the 3FD freeze-out procedure, which is analyzed in
Ref. \cite{3FDfrz} in more detail.

The freeze-out criterion we use is
\begin{eqnarray}
\label{FOcriterion1}
\varepsilon < \varepsilon_{\scr{frz}},% = 0.4 \mbox{GeV/fm}^3,
\end{eqnarray}
where $\varepsilon = u_\mu T^{\mu\nu} u_\nu$
is the total energy density of all three fluids in the proper reference
frame, where the composed matter is at rest. This total energy density is
defined in terms of the total energy--momentum tensor
$T^{\mu\nu} \equiv
T^{\mu\nu}_{\scr p} + T^{\mu\nu}_{\scr t} + T^{\mu\nu}_{\scr f}$
being the sum of energy--momentum tensors $T^{\mu\nu}_{\alpha}$ of
separate fluids (projectile-like, target-like and fireball ones) and
the total collective 4-velocity of the matter
$u^\mu = u_\nu T^{\mu\nu}/u_\lambda T^{\lambda\kappa} u_\kappa$.
Note the latter definition is, in fact, an equation
determining $u^\mu$.
%This definition of the collective 4-velocity is in the spirit of the
%Landau--Lifshitz approach to viscous relativistic hydrodynamics.
%As mentioned above, $\varepsilon_{\scr{frz}}= 0.4$ GeV/fm$^3$ is the
%critical freeze-out energy density.
%
A very important feature of our freeze-out procedure is an anti-bubble
prescription. The matter is allowed to be frozen out  only if \\
{\bf (a)} either the matter is located near the boarder with
vacuum (this piece of matter gets locally frozen out) \\
{\bf (b)} or the maximal value of the total energy density in the system
is less than $\varepsilon_{\scr{frz}}$
\begin{eqnarray}
\label{FOcriterion3}
\max \varepsilon < \varepsilon_{\scr{frz}}
\end{eqnarray}
(the whole system gets instantly frozen out).

In the 3FD model this freeze-out simultaneously terminates both chemical
and kinetic processes.

Before the instant of the global freeze-out, cf. (\ref{FOcriterion3}),
the freeze-out remove matter from the surface of the 
hydrodynamically expanding system.  This removed matter
gives rise to observable spectra of hadrons.
This kind of freeze-out is similar to the model of ``continuous
emission'' proposed in Ref. \cite{Sinyukov02}. There the particle emission
occurs from a surface layer of the mean-free-path width. In our case the
physical pattern is the same, only the mean free path is shrunk to zero.

\begin{figure}[thb]
\includegraphics[width=6.1cm]{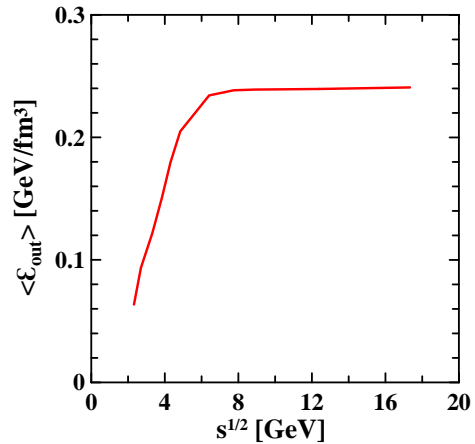}
%\hspace*{-11mm}
%\includegraphics[width=4.0cm]{Nfr(tg).eps}
%$\;$\vspace*{9mm}
\caption{(Color online)
Actual average freeze-out energy density
in central (zero impact parameter) Pb+Pb collisions as a function of
invariant incident energy.
}
\label{fig2a}
\end{figure}

Condition (\ref{FOcriterion1}) 
ensures only that the
actual freeze-out energy density
(let us call it $\varepsilon_{\scr{out}}$),
at which the freeze-out actually occurs,
%$\varepsilon\approx\varepsilon_{\scr{out}}$,
is less than
$\varepsilon_{\scr{frz}}$.
Therefore, $\varepsilon_{\scr{frz}}$ can be called a "trigger" value of
the freeze-out energy density.
As explained in Ref. \cite{3FDfrz},
%Accordingly to Eqs. (\ref{eq1})--(\ref{theta-frz}),
a natural value
of this actual freeze-out energy density is $\varepsilon_{\scr{out}}
\approx \varepsilon_s/2$, i.e. at that the middle of the fall from
the near-surface value of the energy density, $\varepsilon_s$, to
zero. To find out the actual 
value of $\varepsilon_{\scr{out}}$, we have to analyze results of
a particular simulation. In our previous paper \cite{3FD}
we have performed only a rough analysis of this kind.
This is why in the main text of Ref. \cite{3FD} we mentioned the value
of approximately 0.2 GeV/fm$^3$ for $\varepsilon_{\scr{out}}$ and in appendix
explained how the freeze-out actually proceeded. 
%\footnote{
[In terms of
  Ref. \cite{3FD} 
($\varepsilon_{\scr{frz[1]}}$ and 
$\varepsilon_{\scr{frz[1]}}^{\scr{code}}$)
our present quantities are 
$\varepsilon_{\scr{frz}}=\varepsilon_{\scr{frz[1]}}^{\scr{code}}$
and 
$\varepsilon_{\scr{out}}=\varepsilon_{\scr{frz[1]}}$.]
%}.
Results of more comprehensive analysis for central ($b=0$) Pb+Pb
collisions are presented in Fig. \ref{fig2a},
which shows the  $\varepsilon_{\scr{out}}$
value averaged over space--time evolution of the collision:
$\langle\varepsilon_{\scr{out}}\rangle$. As seen,
$\langle\varepsilon_{\scr{out}}\rangle$ reveals saturation at the SPS
energies, very similar to that in effective temperatures in Fig. \ref{fig1}.
This happens in spite of the fact that our freeze-out condition
involves only a single constant parameter $\varepsilon_{\scr{frz}}$.

The "step-like" behavior of $\langle\varepsilon_{\scr{out}}\rangle$
is a consequence of the freeze-out dynamics, as it was demonstrated in Ref.
\cite{3FDfrz}. At low (AGS) incident energies, the energy density
achieved at the border with vacuum, $\varepsilon_s$, is lower than
$\varepsilon_{\scr{frz}}$. Therefore, the surface freeze-out starts
at lower energy densities. It further proceeds at lower densities up
to the global freeze-out because the freeze-out front moves not
faster than with the speed of sound, like any perturbation in the
hydrodynamics. Hence it cannot overcome the
supersonic barrier and reach dense regions inside expanding system.
With the incident energy rise the energy density achieved at the
border with vacuum gradually reaches the value of
$\varepsilon_{\scr{frz}}$ and then even overshoot it. If the
overshoot happens, the system first expands without freeze-out. The
freeze-out starts only when $\varepsilon_s$ drops to the value of
$\varepsilon_{\scr{frz}}$. Then the surface freeze-out occurs really
at the value $\varepsilon_s \approx \varepsilon_{\scr{frz}}$ and
thus the actual freeze-out energy density saturates at the value
$\langle\varepsilon_{\scr{out}}\rangle \approx
\varepsilon_{\scr{frz}}/2$. This freeze-out dynamics is quite stable
with respect to numerics \cite{3FDfrz}.

\begin{figure}[thb]
\includegraphics[width=7.4cm]{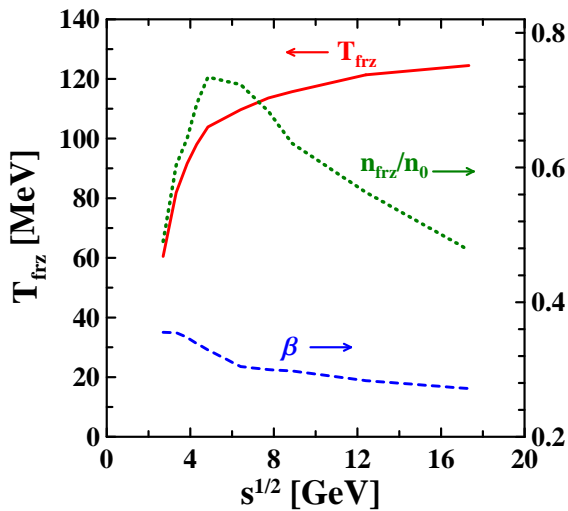}
\caption{(Color online)
Average temperature ($T_{\scr{frz}}$, see scale at the l.h.s.
axis), transverse velocity  ($\beta=v_{T}/c$, see scale at the r.h.s.
axis) and baryon density over the normal nuclear density
($n_{\scr{frz}}/n_0$, see scale at the r.h.s. axis) at the
freeze-out in central  Au+Au (at AGS energies, $b=2$ fm) and Pb+Pb (at
SPS energies, $b=2.5$ fm) collisions as a function of invariant
incident energy. Arrows point to the proper axis scale.
}
\label{fig2b}
\end{figure}

Fig. \ref{fig2b} presents average
temperatures, transverse velocities and baryon densities achieved at the
freeze-out in central collisions. The freeze-out temperature 
$T_{\scr{frz}}$ has a similar "step-like" behavior. However, its
absolute values are essentially lower than effective temperatures in
Fig. \ref{fig1}. This fact once again illustrates that
inverse slopes
represent a combined effect of the temperature and collective
transverse flow of expansion associated with the collective transverse
velocity  $\beta$. Note that $\beta$ even slightly decreases with
incident energy. At SPS energies the freeze-out temperatures in
Fig. \ref{fig2b} are
noticeably lower than those deduced from hadron multiplicities in the
statistical model \cite{Andronic06,Cleymans06}.
The reason for this is as follows. Whereas
the statistical model assumes a single uniform fireball,
in the 3FD simulations at the late stage of the evolution the
system effectively consists of several ``fireballs'':
two (one baryon-rich
and one baryon-free) fireballs at lower SPS energies and three (two
baryon-rich
and one baryon-free) fireballs at top SPS energies \cite{3FDfrz}. 
%At the top SPS energies these
%three fireballs turn out to be even spatially separated. 
Therefore,
whereas high multiplicities of mesons and antibaryons are achieved by
means of high temperatures in the statistical model, the 3FD model
explains them by an additional contribution of the baryon-free
fireball at a lower temperature. In particular, this is the reason why
two different freeze-out points (chemical and kinetic ones) are not
needed in the 3FD model. 
The freeze-out baryon density
$n_{\scr{frz}}$ exhibits a maximum at incident energies of
$E_{\scr{lab}}=$ 10$A$--30$A$ GeV
which are well within range of the planned FAIR in GSI.
This observation agrees with that deduced from the statistical model
\cite{Randrup06}, even  baryon density values in the maximum are
similar to those presented in Ref. \cite{Randrup06}.

Returning to the question if the considered "step-like"
behavior of effective temperatures is a signal of phase transition
into QGP, we should admit that this is not quite clear as yet.
It depends on the nature of the freeze-out parameter
$\varepsilon_{\scr{frz}}=0.4$ GeV/fm$^3$ which should be further
clarified. EoS is not of prime importance for this %"step-like"
behavior. The only constrain on the EoS is that it should be
in some way reasonable. 
Moreover, our preliminary results indicate
that a completely different EoS \cite{Toneev06} with 1st order phase
transition to QGP still reasonably reproduces this "step-like"
behavior even in spite of that it fails to describe a large
body of other data. This happens 
because the same freeze-out pattern is accepted there.

In fact, EoS is just the pressure as
a function of baryon and energy densities: $P(n_B,\varepsilon)$.
In this calculation we used hadronic EoS \cite{3FD,gasEOS}. However,
this is just {\it an interpretation} of the function
$P(n_B,\varepsilon)$, which we use, in hadronic terms. Moreover, our EoS
is too soft
%\footnote{Though the directed flow requires even  softer EoS at top
%AGS and SPS energies \cite{3FDflow}.}
%
at high densities to be matched with even heavy-quark
bag-model EoS \cite{Toneev06} or quasiparticle fits to lattice
QCD data \cite{Ivanov05} in order to construct the 1st order
phase transition.
%\footnote{
[The hadronic and quark pressures as
functions of the baryon chemical potential should have a crossing
point in order to construct the 1st order phase transition.]
%}.
%
Therefore, it would not be surprising
if the same EoS can be reinterpreted also in terms of
very smooth cross-over phase transition to quark--gluon matter.

This hydrodynamic explanation of the considered "step-like"
behavior of effective temperatures together with failure of kinetic
approaches implies that a heavy nuclear system really reveals
a hydrodynamic motion during its expansion.
Simultaneous reproduction of inverse-slopes of all considered
particles ($p$, $\pi$ and $K$) 
implies that these particles belong to 
the same hydrodynamic flow at the instant of their freeze-out. 
An indirect support of this conjecture is the recent success of 
the GiBUU model \cite{Larionov07} in reproduction of kaon
inverse-slopes. That was achieved by taking into account 
three-body interactions, which essentially increased the
equilibration rate.

%\vspace*{5mm} {\bf Acknowledgements} \vspace*{5mm}

We are grateful to I.N. Mishustin, L.M. Satarov,
V.V.~Skokov, V.D. Toneev,  and D.N. Voskresensky for fruitful
discussions.
This work was supported %in part by
the Deutsche
Forschungsgemeinschaft (DFG project 436 RUS 113/558/0-3), the
Russian Foundation for Basic Research (RFBR grant 06-02-04001 NNIO\_a),
Russian Federal Agency for Science and Innovations
(grant NSh-8756.2006.2).

\end{document}